# Estimation of High Impedance Fault Location in Electrical Transmission Lines Using Artificial Neural Networks and R-X Impedance Graph

Serkan BUDAK[1,*] and Bahadır AKBAL[2]

[1,2]Konya Technical University

[1,*]sbudak@ktun.edu.tr, [2]bakbal@ktun.edu.tr

*Abstract*

*It is very important to ensure continuity in the process from generation of electricity to transmission to cities. The most important part of the system is energy transmission lines and distance protection relays that protect these lines. The main function of the protection relays in electrical installations should be deactivated as soon as possible in the event of short circuits in the system. An accurate error location technique is required to make fast and efficient work. Distance relays are widely used as main and backup protection in transmission and distribution lines. Basically, distance protection relays determine the impedance of the line by comparing the voltage and current values. In this study, artificial neural network (ANN) has been used to accurately locate high impedance short circuit faults in 154 kV power transmission lines. The impedance diagram (R-X) of the circuit breaker, current-voltage transformer, overhead transmission line, distance protection relay and distance protection relay has been formed by using simulation program in order to make the study real. The data sets created by recording the image of the change of the impedance occurring at the time of high impedance short circuit fault. The related focal points in the images are given as input to different ANN models and predicted the short circuit faults occurring at different locations on the transmission lines with high accuracy.*

## 1. Introduction

In modern power systems, it is very important to ensure continuity in the process from generation of electricity to transmission to cities. The most important part of the system is energy transmission lines and distance protection relays that protect these lines. The main function of protection relays in electrical installations is to ensure that the faulty zone is taken out of service as soon as possible in the event of short circuits in the system or in any abnormal situation that may damage the facility and the facility components.

Distance protection relays are widely used as main and backup protection in transmission and distribution lines. It is the most preferred relay to protect transmission lines [1]. Basically distance protection relays determine voltage and current values according to line impedance. When the distance protection relays on the transmission and distribution lines encounter a high impedance fault (HIF), they cannot detect short circuit faults in some cases. Most HIF occur when the phase conductor usually comes into contact with wood, poles, road surface, pavement, grass and high-impedance soil [2]. Undetected faults can cause harm to humans and animals, can cause fires, cause power outages, production losses, and damage facility components.

Many studies conducted in the literature on the detection of HIF [2-4]. According to a report in [5], only 35 were detected by the relays from 200 HIF in the five distribution feeders. A detailed analysis of all the techniques proposed so far to detect HIF can be found [6] and studies can be divided into four categories. These are techniques based on impedance measurement, techniques based on traveling-wave, techniques based on high frequency components of current and voltage in the event of a fault, techniques based on knowledge-based approaches [7]. These are techniques based on impedance measurement, techniques based on traveling-wave, techniques based on high frequency components of current and voltage in the component of a fault, techniques based on knowledge-based approaches [7]. Developing technology and increasing needs have increased the range of interest of this basic function of the relay and new protection methods have been developed to ensure correct coordination. In this study, transmission lines are modeled in PSCAD ™ / EMTDC ™ simulation program and the operation of distance protection relay in (HIF) is investigated. As a result of the operation, the fault of the distance protection relay and the location of the fault could not be determined. In this study, a new approach based on artificial neural networks is proposed to detect HIFs.

## 2. Distance Protection Relay in Energy Transmission Lines

Fault detection and fault classification are the two most important elements in the transmission line protection. Basically, distance protection relays determine the voltage and current values according to the impedance of the line. If the impedance value measured by the relay is less than the set value previously entered into the relay, the relay operates.

$$Z=V/I \qquad (1)$$

It works if the measured impedance is less than or equal to the set impedance.

Relay opens if $Z < Z_{set}$

The R-X impedance diagram in Figure 1. shows the distance protection relay to determine whether it is within the protection zone by calculating the impedance value.

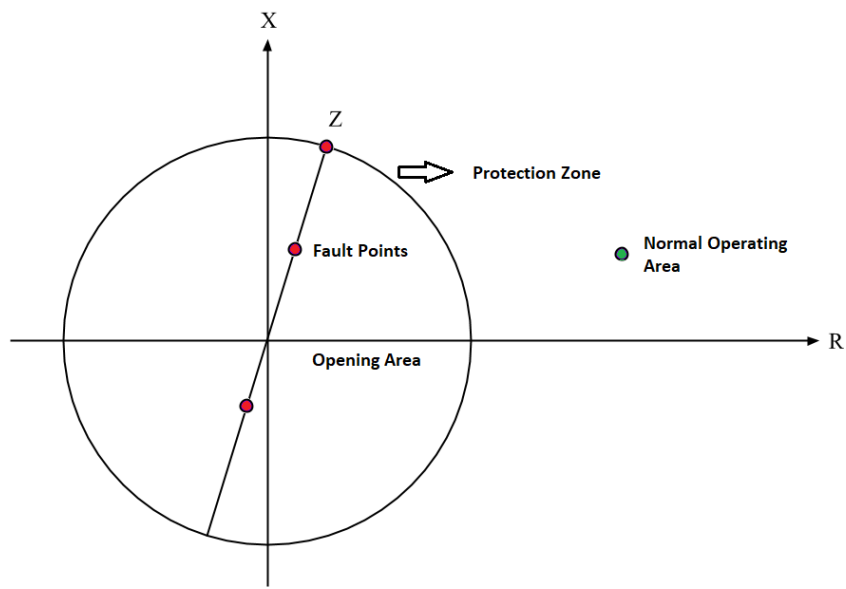

**Figure 1. The Basic Protection Zone of The Distance Protection Relay in The R-X Diagram [6].**

## 3. High Impedance Faults in Energy Transmission Lines

In normal short circuit faults, the distance protection relay calculates the impedance and sends a trip signal to the circuit breaker according to the R-X impedance diagram set as the protection zone and locates the fault. In the event of a HIF, the impedance relays cannot detect the fault and failing this detection causes serious problems. Figure 2. shows that the low-impedance fault enters the protection zone and Figure 3. shows the voltage-current graph that occurs during a short circuit. In Figure 4. high-impedance fault does not enter the protection zone and Figure 5. shows the voltage-current graph that occurs during a short circuit. In case of high impedance faults that do not enter the protection zone, the distance protection relay does not operate correctly.

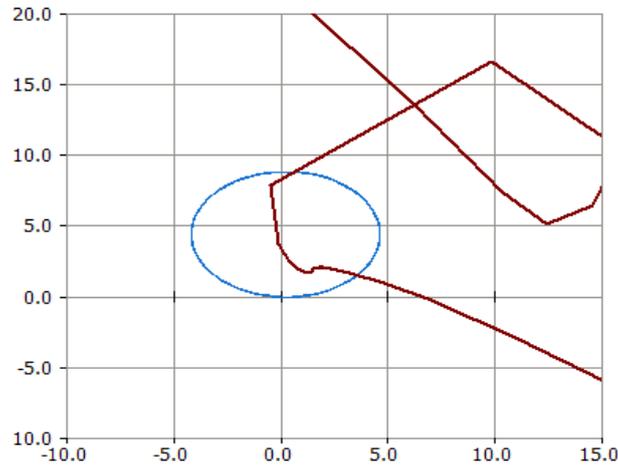

**Figure 2. Low İmpedance Short Circuit Fault İnstantly İmpedance Diagram**

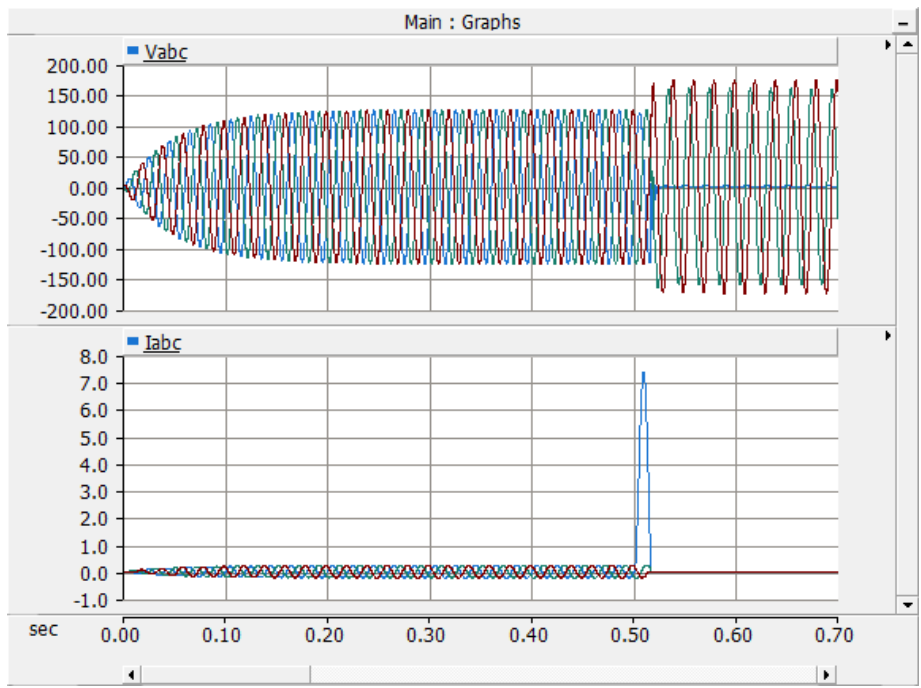

**Figure 3. Low İmpedance Short Circuit Fault İnstantly Voltage-Current Graph**

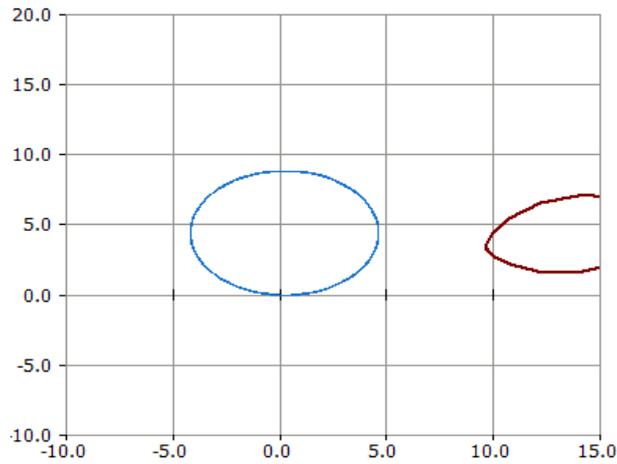

**Figure 4. High İmpedance Short Circuit Fault İnstantly İmpedance Diagram**

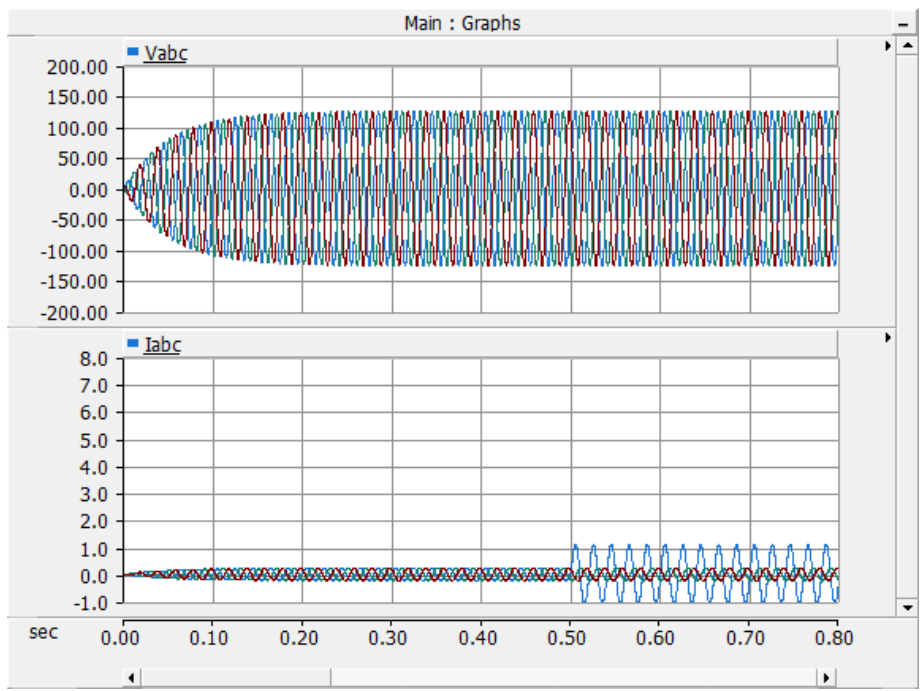

**Figure 5. High Impedance Short Circuit Fault İnstantly Voltage-Current Graph**

## 4. Fault and Location Detection by Artificial Neural Networks

In the PSCAD ™ / EMTDC ™ simulation program, Figure 6. is modeled as 154kV, 50Hz, double-feed transmission line. Using current and voltage transformers, breaker and distance protection relay in the system, the operation of distance protection relay in different fault types is examined.

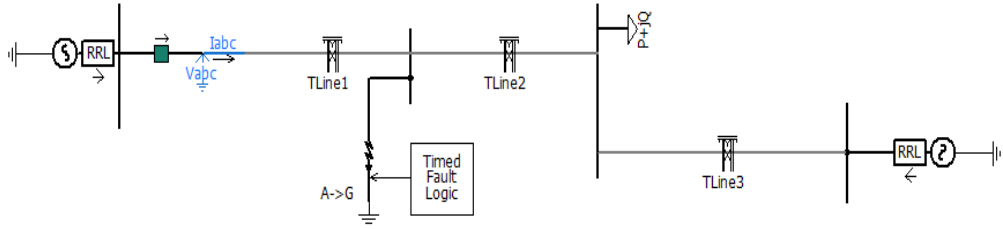

**Figure 6. Transmission Line Model**

In the simulation program, the faults at different impedance affect the operation of the distance protection relay. In this study, the images taken from the R-X impedance diagram of different fault impedances transfer to MATLAB program and image processing methods were apply to ANN program. Phase (a) -ground faults generate at 5, 10, 15, 20, 25, 30, 35, 40, 45 and 50 km of the overhead power transmission line simulated using PSCAD ™ / EMTDC ™. Fault simulations performance using 50 ohm and 100 ohm fault resistors in each of these kms. In this study, data set file was prepared with images taken from R-X impedance diagram. As the ANN inputs, the images taken from the R-X impedance diagram from the distance protection relay are used. When the evaluations related to normalization studies in the literature examine as ANN outputs, 'D Min Max Normalization' method is found to be appropriate [8]. The data set found to be good in the normalization scale of (0.1 - 0.9). With this data set, various network topologies, learning strategies and performance functions tested and the most suitable network structure has been determined. Fault and fault location problems for phase (a) - ground faults are obtained by using forward-feed back propagation ANN and gradient descent momentum and an adaptive learning rate (GDX), scaled conjugate gradient method (SCG) and conjugate gradient backpropagation with Powell-Beale restarts (GDM) training function. Figure 7. shows the regression fit curves with Matlab / nntool output, which include training, validation and test groups together.

**Table 1. Phase (a) - Ground Faults Estimated (km) Fault Locations When Fault Resistance is 50 Ohms**

| Real Distance(km) | TRAINGDX | TRAINSCG | TRAINCGB |
|---|---|---|---|
| 5 | 5.0047 | 5.6884 | 5.0150 |
| 10 | 9.9945 | 9.0943 | 9.9619 |
| 15 | 14.9964 | 15.0256 | 15.4637 |
| 20 | 19.9982 | 20.0371 | 20.0171 |
| 25 | 25.0032 | 24.9696 | 24.7611 |
| 30 | 30.0012 | 29.9883 | 30.0542 |
| 35 | 35.0123 | 35.1008 | 35.0303 |
| 40 | 40.0010 | 39.9837 | 40.0035 |
| 45 | 44.9976 | 45.0961 | 45.4221 |
| 50 | 50.3710 | 49.9445 | 50.0054 |
| **MSE** | 9.78068E-07 | 9.36063E-06 | 3.2351E-06 |

Faults have been generated at various fault locations for fault location detection of the fault resistance and the pre-trained network has been tested with inputs for these faults. According to the 50 ohm fault resistance test results given in Table 1. the highest mean square error 3.2351E-06, the lowest mean square error is 9.78068E-07. According to these results, it is seen that the accuracy of the algorithm is independent of high fault resistance and TRAINGDX training function gives the best results. Figure 7. shows the training, validation and test regression fit curves of the TRAINGDX training function, which gives the best estimate.

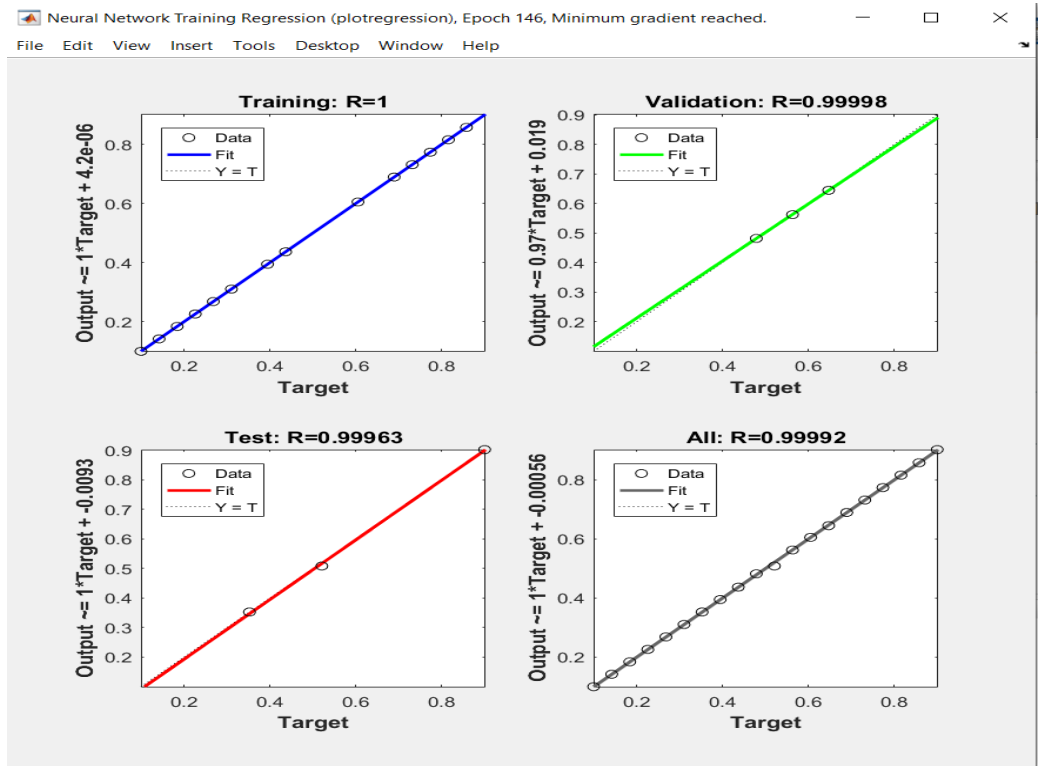

**Figure 7. Training, Validation and Test Regression Fit Curves of The TRAINGDX Training Function That Gives The Best Estimate.**

**Table 2. Phase (a) - Ground Faults Estimated (km) Fault Locations When Fault Resistance is 100 Ohms**

| Gerçek Mesafe(km) | TRAINGDX | TRAINSCG | TRAINCGB |
|---|---|---|---|
| 5 | 6.3033 | 4.8735 | 5.0150 |
| 10 | 9.2603 | 9.0365 | 9.9619 |
| 15 | 14.8303 | 14.5341 | 15.4637 |
| 20 | 19.9500 | 19.2353 | 20.0171 |
| 25 | 24.8607 | 27.5460 | 24.7611 |
| 30 | 29.8627 | 28.9111 | 30.0542 |
| 35 | 34.8649 | 33.1401 | 35.0303 |
| 40 | 39.8823 | 39.4266 | 40.0035 |
| 45 | 44.9419 | 43.9856 | 45.4221 |
| 50 | 49.8195 | 50.4704 | 50.0054 |
| MSE | 1.69019E-05 | 1.02483E-04 | 3,2351E-06 |

Faults have been generated at various fault locations for fault location detection of the fault resistance and the pre-trained network has been tested with inputs for these faults. According to the test results given in 100 ohm fault resistance in Table 2. the highest mean square error is 1.02483E-04, the lowest mean square error is 3,2351E-06. According to these results, it is seen that the accuracy of the algorithm is independent of high fault resistance and TRAINCGB training function gives the best results. Figure 8.

shows the training, validation and test regression fit curves of the TRAINCGB training function, which gives the best estimate.

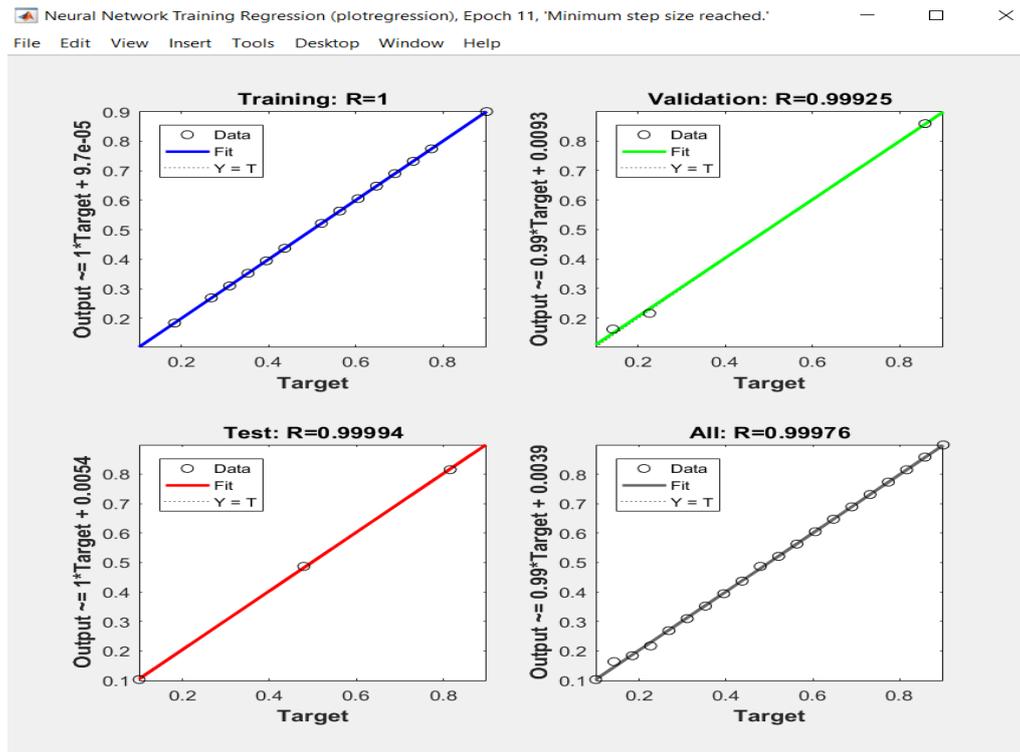

**Figure 8. Training, Validation and Test Regression Fit Curves of The TRAINCGB Training Function Giving The Best Estimate.**

## 5. Conclusions

In this study, an ANN based intelligent system has been designed to accurately identify high impedance short circuit faults and fault location in high voltage power transmission lines. In this new system, images taken from the R-X impedance diagram from the distance protection relay use as ANN inputs. The study was tested in a system modeled with PSCAD ™ / EMTDC ™.

All neural network structures have been tried to be used in the fault and fault locating algorithm based on ANN use and the three highest training algorithms used. When Table 1. and Table 2. are examined, it seen that the errors obtained in the fault and fault location determination study very low and the algorithm shows a good performance. The distance of the defect distances determined is visible in overhead lines.

The advantage of this study over other studies is the design of an ANN based intelligent system by processing the images we receive from the distance protection relay without additional cost, additional device and mixed mathematical equations.